\documentclass[preprint,secnumarabic,amssymb, amsmath,bibnotes, aps, prb, superscriptaddress,floatfix,showpacs]{revtex4-1}
\usepackage{graphicx}
\usepackage{subcaption}
\usepackage{mathrsfs}
\usepackage{hyperref}
\usepackage{multirow}

\newcommand{\degSign}{\ensuremath{^\circ } }
\newcommand{\degSignNoSpace}{\ensuremath{^\circ}}
\begin{document}

\title{de Haas-van Alphen measurement of the antiferromagnet URhIn$_5$}%

\author{Jing Fei Yu}%
\email[email: ]{jfeiyu@physics.utoronto.ca}
\affiliation{Department of Physics, University of Toronto, Toronto, Ontario, M5S 1A7, Canada}
\author{Attila Bartha}
\author{Jeroen Custers}
\affiliation{Department of Condensed Matter Physics, Charles University, Ke Karlovu 5, 121 16 Praha 2, Czech Republic}
\author{Stephen R. Julian}
\affiliation{Department of Physics, University of Toronto, Toronto, Ontario, M5S 1A7, Canada}
\affiliation{Canadian Institute for Advanced Research, 180 Dundas St.W, Toronto, Ontario, M5S 1Z8, Canada}

\date{\today}
\begin{abstract}
    We report on the results of a de Haas-van Alphen (dHvA) measurement performed on the recently discovered antiferromagnet URhIn$_5$ ($T_N$ = 98 K), a 5\textit{f}-analogue of the well studied heavy fermion antiferromagnet CeRhIn$_5$. The Fermi surface is found to consist of four surfaces: a roughly spherical pocket $\beta$, with $F_\beta \simeq 0.3$ kT; a pillow-shaped closed surface, $\alpha$, with $F_\alpha \simeq 1.1$ kT; and
    two higher frequencies $\gamma_1$ with $F_{\gamma_1} \simeq 3.2$ kT and $\gamma_2$ with $F_{\gamma_2} \simeq 3.5$ kT that are seen only near the \textit{c}-axis, and that may arise on cylindrical Fermi surfaces. The measured cyclotron masses range from 1.9 $m_e$ to 4.3 $m_e$. A simple LDA+SO calculation performed for the paramagnetic ground state shows a very different Fermi surface topology, demonstrating a need for more advanced electronic structure calculations. 

\end{abstract}

\maketitle
\section{Introduction}
Uranium based compounds exhibits a wide range of exotic properties, from the unconventional superconductivity in heavy fermion UPt$_3$ \cite{McMullan2008} to the enigmatic hidden order in URu$_2$Si$_2$\cite{Mydosh2014b}, and all the way to weakly-correlated metallic behaviour in the uranium metal itself \cite{Mydosh2014b}. U atoms possess 5\textit{f} electrons, whose spatial extent is greater than that of the 4\textit{f} valence orbitals of the rare earths, but less than that of 3\textit{d}
orbitals of the row four transition metals \cite{Bartha2015}. Depending on their environment, the 5\textit{f} electrons can be localized or itinerant (\textit{i.e.} they may or may not be included in the Fermi volume) \cite{Zwicknagl2003}. Moreover, uranium can have 5\textit{f} valence ranging from 5\textit{f}$^{3}$ to 5\textit{f}$^{0}$, presenting further opportunities for inter- as well as intra-orbital correlations and complexity. 


For some uranium compounds such as the heavy fermion UPt$_3$ \cite{McMullan2008}, the Fermi surface has been mapped and the 5\textit{f} electrons are understood to be itinerant in character. On the other hand, the 5\textit{f} electrons are fully localized in the related compound UPd$_3$ \cite{Tokiwa2001b}, while quantum oscillation data on the heavy fermion superconductor UPd$_2$Al$_3$ \cite{Inada1999} has been explained by band structure calculations in which some
5\textit{f} electrons are localized on the U site, while others are itinerant band electrons \cite{Zwicknagl2003}. Whether and when the 5\textit{f} electrons are localized or itinerant, or partially localized, is still not well understood at present.

The focus of this paper is URhIn$_5$, a member of the U$_n$\textit{TX}$_{3n+2}$ (n = 0,1,2; \textit{T} = transition metal; \textit{X} = In, Ga) compounds. They are isostructural with the Ce$_n$\textit{TX}$_{3n+2}$ family \cite{Bartha2015}. In addition to the open question of the character of the 5\textit{f} electrons, these compounds are also generally interesting because the system has a layered tetragonal structure, which means that adding
layers of \textit{TX}$_2$ can change the dimensionality from 3D to 2D \cite{Bartha2015}.  

URhIn$_5$ has been well characterized along with its bilayer cousin, U$_2$RhIn$_8$ \cite{Bartha2015,Matsumoto2014}. It is tetragonal with HoCoGa$_5$-type structure (P4/mmm). It orders antiferromagnetically at a N\'eel temperature of $T_N = 98$ K \cite{Bartha2015}. The Sommerfeld coefficient is given as $\gamma = 60.7$ mJ mol$^{-1}$K$^{-2}$
with Debye temperature $\Theta_D = 165$ K in Ref. \onlinecite{Bartha2015}, whereas $\gamma = 50$ mJ mol$^{-1}$K$^{-2}$, $\Theta_D = 187$ K are reported by Ref. \onlinecite{Matsumoto2013}. The Sommerfeld coefficient is similar to that of UNiGa$_5$ \cite{Tokiwa2001}, and rather large considering the high N\'eel temperature. The magnetic moment, $\mu_{\rm eff} = 3.6 \mu_B/U$ \cite{Bartha2015}, is close to the U$^{4+}$ (5\textit{f}$^2$) moment
  of 3.58 $\mu_B/U$ \cite{Tokiwa2001}. These are again properties very similar to
 UNiGa$_5$ \cite{Tokiwa2001}. 

 Zero field NMR and NQR measurements have been performed on URhIn$_5$ \cite{Sakai2013}. The nuclear relaxation rate $1/T_1T$ satisfies the Korringa relation just below $T^* \approx 150$ K, which has led to the suggestion that the 5\textit{f} electrons are localized above $T^*$ and itinerant below \cite{Sakai2013}. The AFM propagation vector is commensurate, with propagation vector $\mathbf{Q} = (1/2,1/2,1/2)$ as determined unambiguously by recent neutron diffraction
 study \cite{Bartha2017}. Interestingly, UNiGa$_5$ also exhibits the same ordering vector \cite{Sakai2013}.

Matsumoto \textit{et al.} \cite{Matsumoto2014} have previously reported dHvA quantum oscillations in URhIn$_5$. They only observed two frequencies, 1.1 kT and 720 T along [100], with angle dependence suggesting that they both arise on a small ellipsoidal pocket. The presence of small pockets is consistent with the non-magnetic, 5\textit{f} itinerant semi-metal URhGa$_5$ \cite{Ikeda2002}, but the authors note that the large specific heat for the antiferromagnetic state of URhIn$_5$ cannot
be explained by this pocket alone, suggesting the likely existence of larger Fermi surfaces undetected by their study. In comparison, quantum oscillation measurements on UNiGa$_5$ show one cylindrical surface (1.3 kT) along with two larger ellipsoidal surfaces (2.6 kT and 1.6 kT), but additional heavier Fermi surfaces would again be needed to explain the large specific heat \cite{Tokiwa2001}. On the other hand, a larger Fermi surface (4 kT) has been detected, in addition to smaller pockets, in UIn$_3$, the $n=\infty$ member of the  
U$_n$\textit{TX}$_{3n+2}$ family, although additional, multi-connected surfaces are still needed for $\gamma$ to match experimental values \cite{Tokiwa2001a}. 

Recently, magnetoresistance and Hall resistance also have been measured on URhIn$_5$ \cite{Haga2017}. The magnetoresistance follows the $H^2$ behaviour at higher temperatures but deviates from it for temperatures below 30 K \cite{Haga2017}. At low temperature the magnetoresistance is positive, but the angle dependence, which could reveal the presence of open orbits, has not been reported. The Hall resistance $R_H$ is negative below the N\'eel temperature ($T_N$),
suggesting that the electron-like carriers dominate electronic transport \cite{Haga2017}. 

Motivated by these previous measurements, we performed a quantum oscillation measurement on high quality URhIn$_5$ samples. In this paper, we present these results in detail. 


\begin{figure}
    \centering
    \label{fig:crystalStructure}
    \includegraphics[scale=0.3]{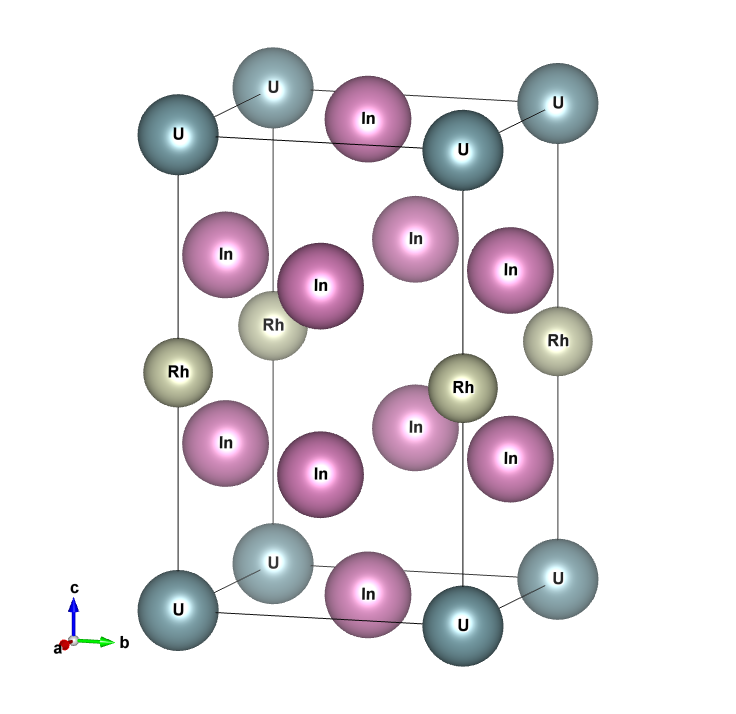}
    \caption{Crystal structure of URhIn$_5$, visualized using VESTA software \cite{Momma2011}}
\end{figure}
\section{Samples and Experimental Method}
Three high quality crystals of URhIn$_5$, prepared using the self-flux method \cite{Bartha2015} were used. The samples were X-rayed and characterized using specific heat before the experiment was performed, and they had extremely high RRR (exceeding 200) \cite{Bartha2015}, ideal for quantum oscillation measurements. 

Modulation field de Haas-van Alphen measurements were carried out on these samples, with field ranges from 9-16 T and at temperatures between 65-2000 mK. The samples were rotated with respect to the applied field in the [001] to [100] plane and the [100] to [110] plane.  

\section{Results}
\begin{figure}
    \centering
    \label{fig:raw_data_and_FFT}
    \begin{subfigure}[t]{\linewidth}
        \centering
        \includegraphics[height=0.25\textheight]{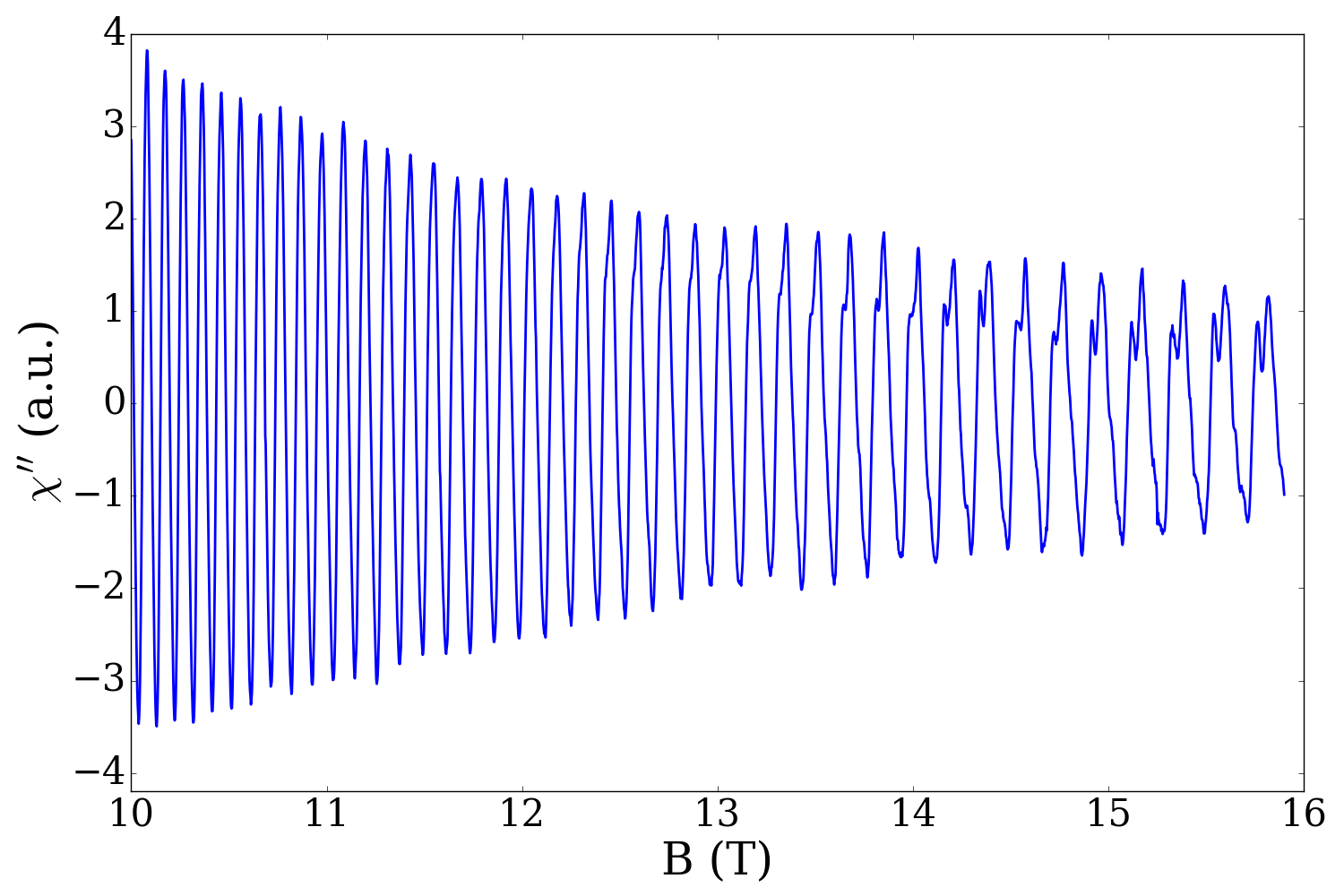}
        \caption{\label{fig:raw_data}}
    \end{subfigure}
    \begin{subfigure}[l]{\linewidth}
        \centering
        \includegraphics[height=0.25\textheight]{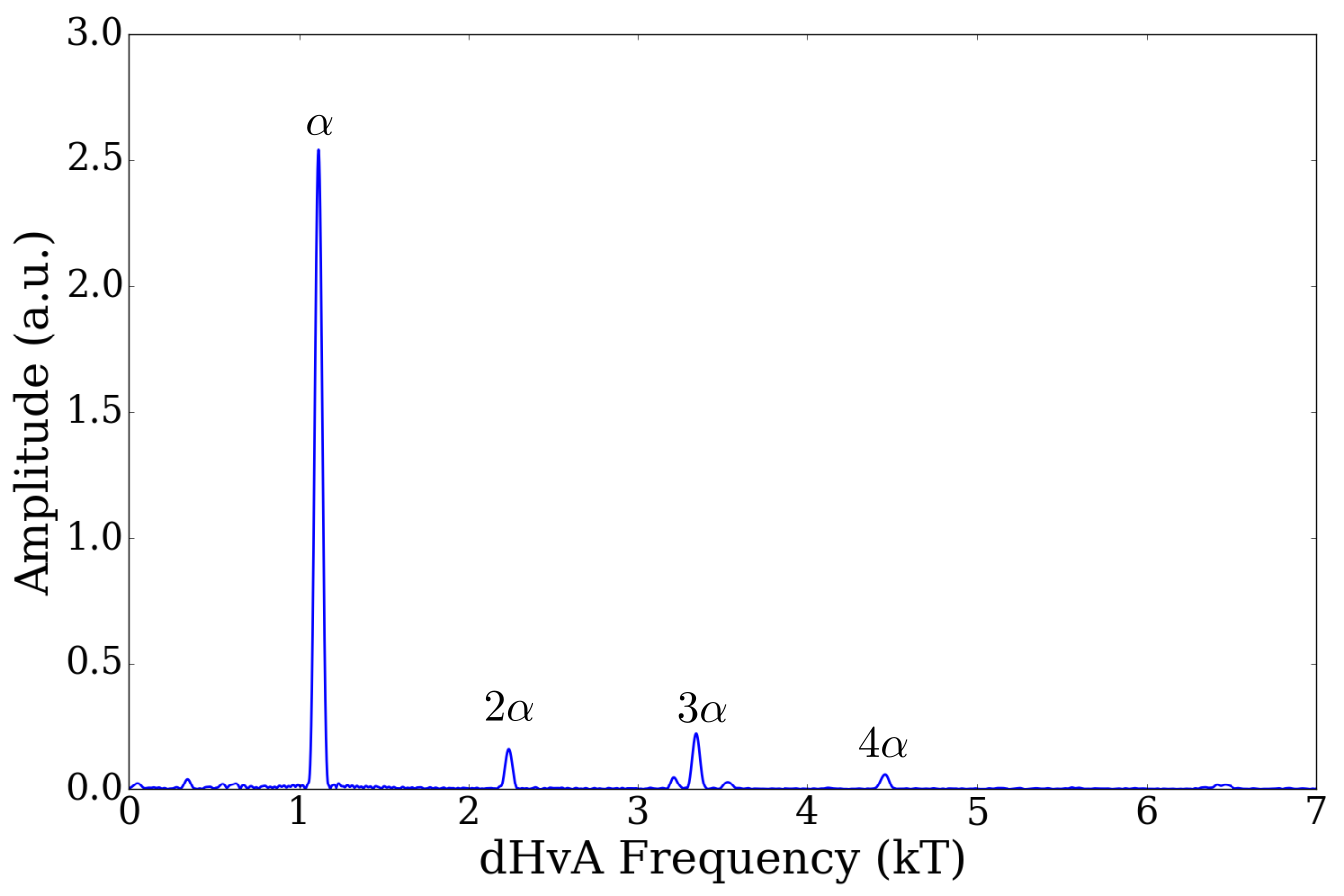}
        \caption{\label{fig:FFT_raw_data}}
    \end{subfigure}
   \begin{subfigure}[l]{\linewidth}
       \centering
       \includegraphics[height=0.25\textheight]{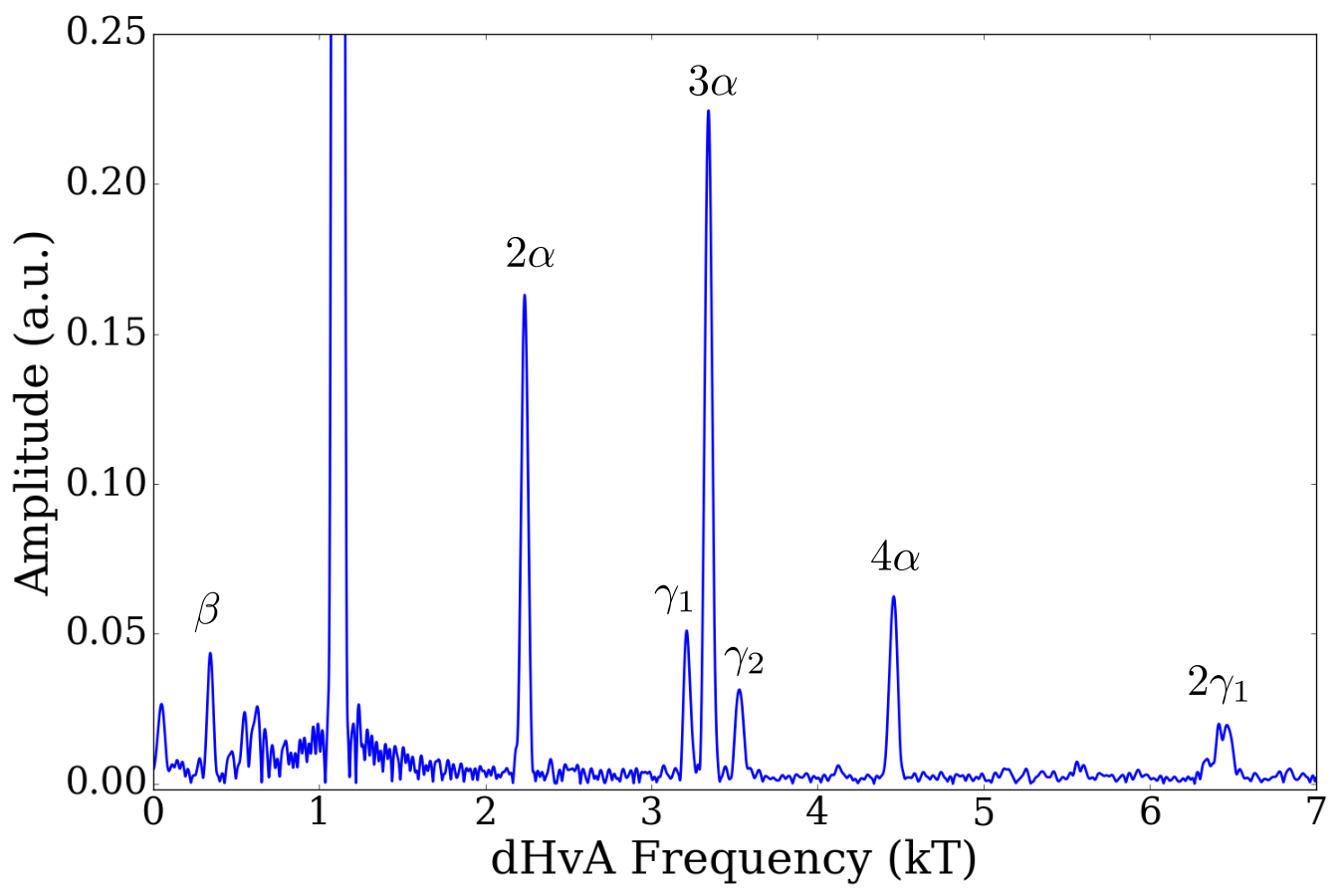}
       \caption{\label{fig:FFT_raw_data_zoomed_in}}
    \end{subfigure}
    \caption{\textbf{\subref{fig:raw_data}}:URhIn$_5$ quantum oscillations from 10 to 16 T, 65 mK. The field is 4\degSign from [001]. \textbf{\subref{fig:FFT_raw_data}}: Fourier transform of the data in (\subref{fig:raw_data}), showing a dominant $\alpha$ frequency that has been previously reported \cite{Matsumoto2014}.  \textbf{\subref{fig:FFT_raw_data_zoomed_in}}: The frequency spectrum of (\subref{fig:FFT_raw_data}) with an expanded vertical scale, revealing several
    new frequencies, not observed in previous work.}
\end{figure}

Fig. \ref{fig:raw_data} shows an example of our raw data from 10 to 16 T at 65 mK, with field tilted at 4\degSign from [001], and Fig. \ref{fig:FFT_raw_data} shows the Fourier transform of the same set of raw data. A very strong peak at 1.1 kT and its three harmonics can be easily seen. This frequency is labelled $\alpha$. In addition, expending the vertical scale in Fig. \ref{fig:FFT_raw_data_zoomed_in}, other sharp peaks are present as well: at 0.3 kT, 3.2 kT and 3.5 kT. These are labelled
$\beta$, $\gamma_1$ and $\gamma_2$ respectively.

\begin{figure}
    \centering
    \includegraphics[width=\linewidth,keepaspectratio]{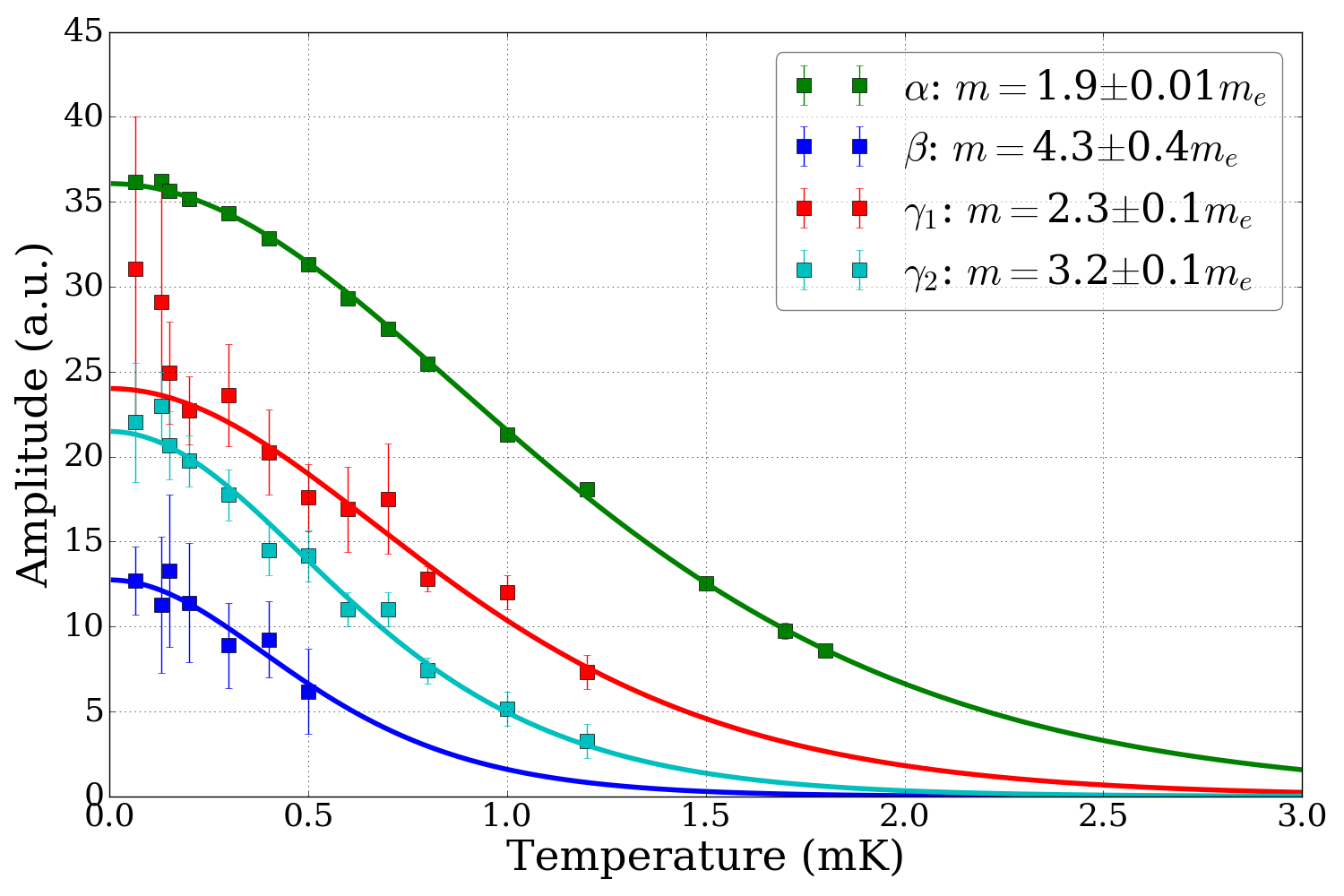}
    \caption{Mass study of different dHvA frequencies with $B\parallel$ [001], using LK fit. The error bars on the $\alpha$ branch are too small to be seen.}
    \label{fig:mass_study}
\end{figure}
The effective mass $m^*$ associated with each extremal orbit can be extracted using the Lifshitz-Kosevich (LK) formula:
\begin{equation}
    A = \frac{Cm^*T/B}{\sinh Cm^*T/B}
    \label{eq:LK}
\end{equation}
where $C=\frac{2\pi^2k_Bm_e}{e\hbar}$. Typical fits are shown in Fig. \ref{fig:mass_study}. The cyclotron masses
found for other directions are summarized in Table \ref{tab:experiment_summary}. 

\begin{table}
    \centering
    \begin{tabular}{|c|cc|cc|cc|}
        \hline
         & \multicolumn{2}{|c|}{B $\parallel$ [100]}&\multicolumn{2}{|c|}{B $\parallel$ [110]}&\multicolumn{2}{|c|}{B $\parallel$ [001]}\\
        \hline
        Orbit & $F$ (kT) & $m^* (m_e)$ & $F$ (kT) & $m^*(m_e)$ & $F$ (kT) & $m^*(m_e)$\\
        \hline
        $\beta$ &    &                &      &             & 0.3 & 4.3$\pm$0.4\\
        $\alpha$ & 1.1 & 1.5$\pm$0.02 & 1.1 & 2.0$\pm$0.2 & 1.1 & 1.9$\pm$0.01\\
        $\gamma_1$&     &               &      &              & 3.2 & 2.3$\pm$0.1\\
        $\gamma_2$&     &               &      &              & 3.5 & 3.2$\pm$0.1\\
        \hline
    \end{tabular}
    \caption{Summary of the experimental dHvA frequencies and fitted cyclotron masses $m^*$. Missing entries mean that the oscillation were only observed with the applied field along the \textit{c}-axis.}
    \label{tab:experiment_summary}
\end{table}

\begin{figure*}
    \centering
    \includegraphics[width=\linewidth,keepaspectratio]{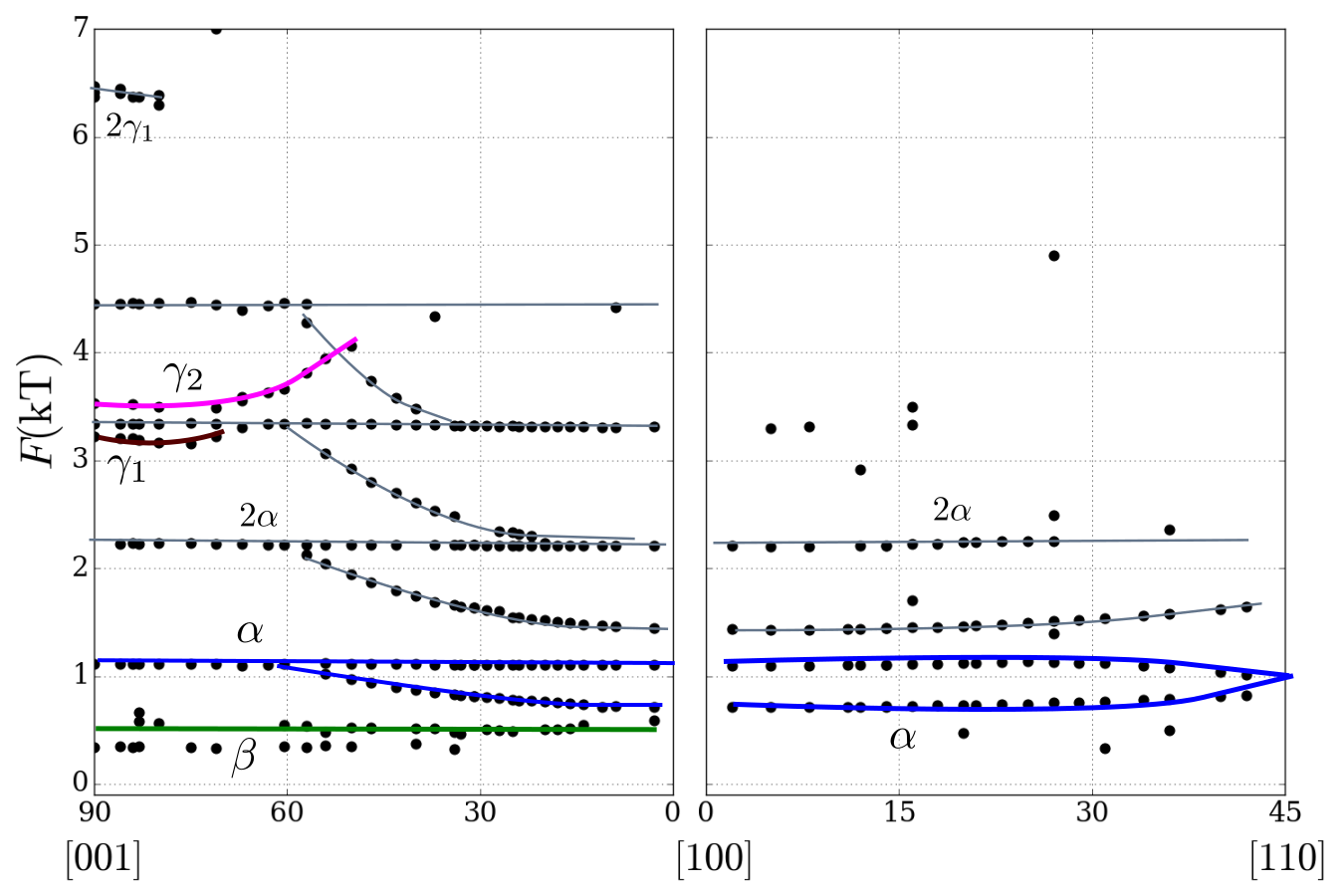}
    \caption{dHvA frequency $F$ vs $\theta$, where $\theta$ is the angle between the magnetic field and the labelled crystalline axes. The data from [001] to [110] plane has been omitted as it is identical to that of [001] to [100]. } 
    \label{fig:combined_rotation}
\end{figure*}
The angle dependence of the dHvA frequencies is shown in Fig. \ref{fig:combined_rotation}. This angle dependence suggests that the $\beta$ branch, which has very little angle dependence, arises from a nearly spherical Fermi surface. However, the lack of any significant angle dependence may also be a sign of an impurity phase from, say, the inclusion of flux used to grow the sample, or perhaps a spherical pocket in an impurity phase such U$_2$RhIn$_8$. The $\alpha$ branch splits into two
frequencies around 40\degSignNoSpace. This suggest a distorted sphere that has only one extremal orbit at $c$-axis, but two extremal orbits beyond a critical angle. The two branches $\gamma_1$ and $\gamma_2$ are close
in frequency and it is natural to think that they arise on the same Fermi surface sheet, geometrically this would be certain only if they converge as the angle is rotated away from the \textit{c}-axis. But the frequency of $\gamma_1$ shows only a small upward curvature before it disappears. Thus we cannot rule out the possibility that they are distinct surfaces. $\gamma_1$ and $\gamma_2$ may arise from cylindrical surfaces, as shown from their upward curvatures. This is supported by the fact that we do not observe
these frequencies at [100] or [110], which we
would expect if this were a closed surface like $\alpha$. However, as $\gamma_1$ and $\gamma_2$
disappear relatively quickly when we rotate away from the \textit{c}-axis, the evidence is not conclusive. 

\begin{table}
    \centering
    \begin{tabular}{|c|c|c|c|}
        \hline
        \multirow{9}{*}{Band 47}& $B$ Direction & $F$(kT) &$m^* (m_e)$\\
        \cline{2-4}
        &\multirow{2}{*}{ [100] }& 1.1034 & 0.73\\
        && 1.1516 &0.96\\
        \cline{2-4}
        &\multirow{1}{*} { [110] } & 0.1309 &0.83 \\
        \cline{2-4}
        &\multirow{4}{*}{ [001] }& 0.2110 & 0.47\\
        && 0.2117 &1.02\\
        && 0.3034 & 0.44\\
        && 0.5469 & 0.65\\
        \hline
        \multirow{4}{*}{Band 48}&\multirow{2}{*}{ [100] }& 1.0037 &1.37\\
        && 1.5812 & 1.42\\
        \cline{2-4}
        &\multirow{1}{*} { [110] } & 1.1677 &1.45\\
        \cline{2-4}
        &\multirow{1}{*}{ [001] }& 1.0745 &2.16\\
        \hline

    \end{tabular}
    \caption{dHvA frequencies found using SKEAF, for $B\parallel$[100], $B\parallel$[110] and $B\parallel$[001].}

    \label{tab:SKEAF_results}
\end{table}

\begin{figure}
    \centering
    \includegraphics[scale=0.4]{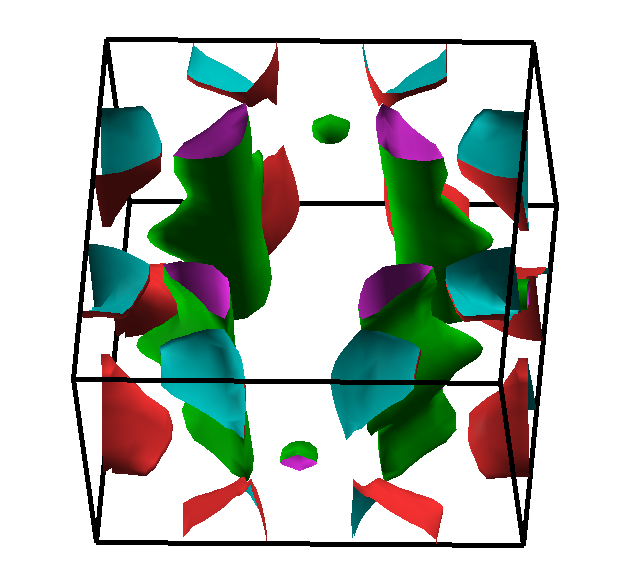}
    \caption{
        Fermi surface of the paramagnetic state in a single BZ, calculated by LDA+SO methods using the WIEN2k package \cite{Wien2k}. Green: band 47; Red: band 48. }
   \label{fig:FS_WIEN2K} 
\end{figure}
\section{Analysis and Discussion}
\subsection{Paramagnetic Ground State Calculation}
We performed an LDA+SO calculation of URhIn$_5$'s ground state, using the linearized augmented plane wave (LAPW) scheme that was implemented in the WIEN2k package \cite{Wien2k}. Spin-orbit coupling was turned on, and RK$_{\mathrm{max}}$ was set to 7.0. The resulting Fermi surface is shown in Fig. \ref{fig:FS_WIEN2K}. Using the Supercell K-space Extremal Area Finder (SKEAF) \cite{SKEAF} program, dHvA frequencies and approximate effective masses were extracted and are given in Table
\ref{tab:SKEAF_results}. These values do not generally match the experimentally obtained values in Table \ref{tab:experiment_summary}, however, band 48 does have a 1.1 kT frequency that is similar to our $\alpha$ branch.  

\begin{figure}
    \centering
    \includegraphics[width=\linewidth,keepaspectratio]{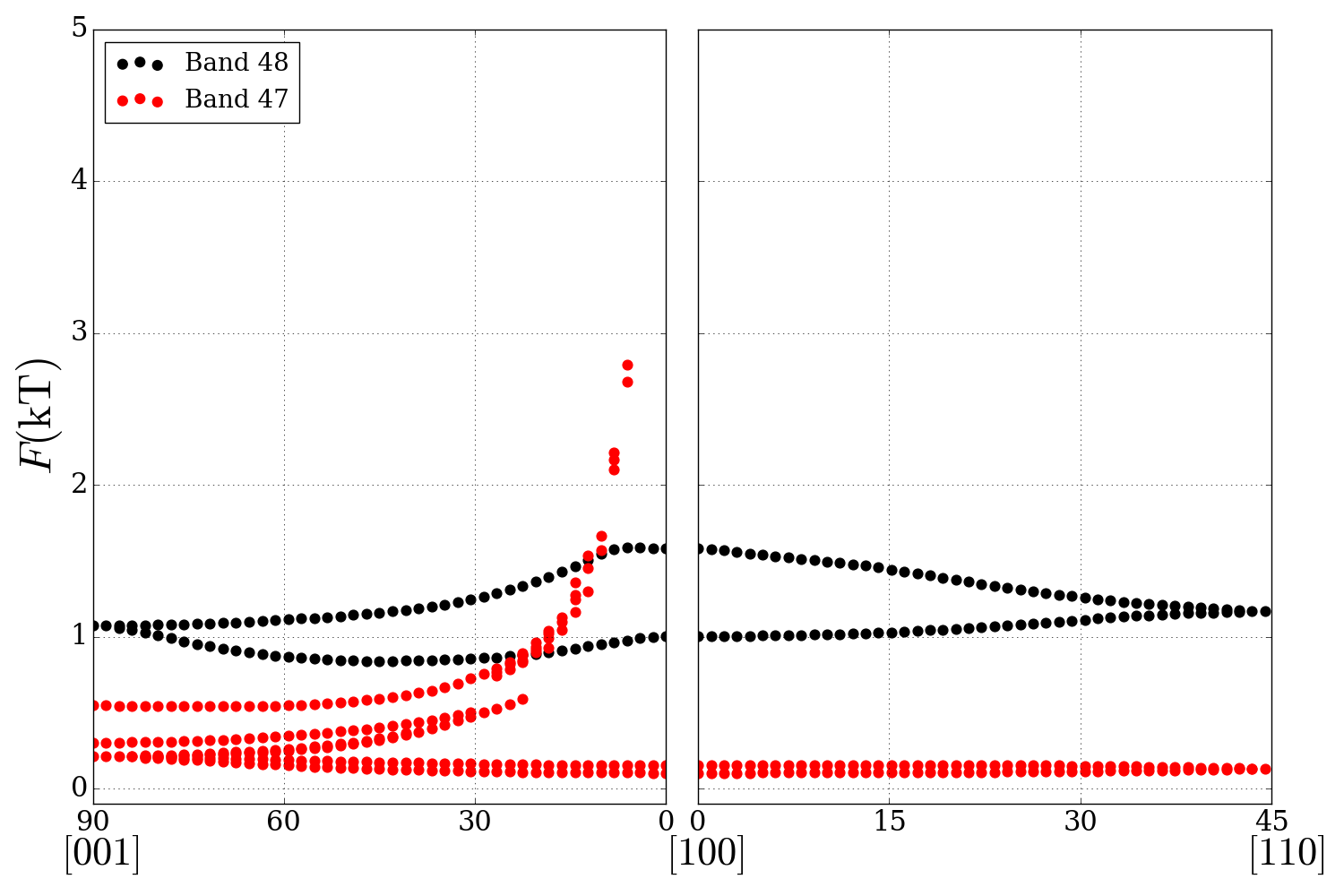}
    \caption{dHvA frequency $F$ vs $\theta$ found by SKEAF for the paramagnetic all-itinerant Fermi surface, where $\theta$ is the angle between the magnetic field and the labelled crystalline axes. For the corresponding surfaces to each band see Fig. \ref{fig:FS_WIEN2K} and its caption.} 
    \label{fig:SKEAF_rotation}
\end{figure}

To study this 1.1 kT frequency further, we used SKEAF to find the angle dependence of the frequencies, shown in Fig. \ref{fig:SKEAF_rotation}. Here we can see that the 1.1 kT frequency is a closed ellipsoidal surface similar to our $\alpha$ branch. However, the difference from experiment is that a second, \textit{higher} frequency of 1.5 kT branches out from the main frequency whereas in experimental results, a \textit{lower} frequency branches out. The topology of the two Fermi surfaces are
therefore not very similar. 
While the $\gamma_1$ and $\gamma_2$ frequencies in our experiment could be coming from cylindrical surfaces,they are almost 10 times larger than the frequencies found by SKEAF in the paramagnetic WIEN2k calculation. It is extremely unlikely that antiferromagnetism could produce larger Fermi surfaces, so our conclusion is that the paramagnetic, all-itinerant Fermi surface is not a good starting point for explaining URhIn$_5$. 


\subsection{Fermi Volume Estimation}
The total Brillouin zone (BZ) volume is $1.6\times10^{30}$ m$^{-3}$. If antiferromagnetism doubles the unit cell in the \textit{c}-axis, the AFM BZ shrinks to $7.8\times10^{29}$ m$^{-3}$. If we assume all the Fermi surface sheets are spherical so the volume of each sheet is given by $V = 4/3\pi k_F^3$, then given the antiferromagnetic BZ, $\alpha$ occupies 3.2\% of the BZ; $\beta$ occupies 0.4\%; $\gamma_1$ occupies 16\% and $\gamma_2$ occupies 19\% of the Brillouin zone.
If we assume $\gamma_1$ and $\gamma_2$ are cylindrical surfaces and their volumes are given by $S_F\times2\pi/(2c)$ where $S_F$ is the area enclosed by the cylinder and given by the dHvA frequency, c is the \textit{c}-axis lattice parameter, and $2\pi/(2c)$ is the height of the AFM BZ, the volumes of $\gamma_1$ and $\gamma_2$ become 17\% and 18\% of the total BZ volume, respectively. In total these Fermi surface sheets occupy nearly 40\% of the total AFM BZ, if we assume there is only one
copy of each Fermi surface in the BZ. As noted in the next section, however, it is probable that there are multiple copies of at least some Fermi surfaces. 

\subsection{Sommerfeld Coefficient Estimation}
We can estimate the Sommerfeld coefficient $C/T = \gamma$ from the Fermi surfaces obtained. If we make the simplifying assumption that the frequencies at [001] correspond to spherical Fermi surfaces, so that the extremal area is given by $S_F=\pi k_F^2$, then $\alpha$ would contribute 1.7 mJ mol$^{-1}$K$^{-2}$; $\beta$ would contribute 2.1 mJ mol$^{-1}$K$^{-2}$; $\gamma_1$ would contribute 3.6 mJ mol$^{-1}$K$^{-2}$ and $\gamma_2$ would contribute 5.2 mJ mol$^{-1}$K$^{-2}$. This sums up to 12.6 mJ mol$^{-1}$K$^{-2}$
 if we assume one sheet each, whereas the experimental
value is around 50-60 mJ mol$^{-1}$K$^{-2}$ \cite{Bartha2015,Matsumoto2013}. 

If, however, we assume that $\gamma_1$ and $\gamma_2$ are distinct, cylindrical, Fermi surfaces. Then their specific heat contributions are no longer $k_F$ dependent and they rise to 7.6 mJ mol$^{-1}$K$^{-2}$ for $\gamma_1$ and 10.6 mJ mol$^{-1}$K$^{-2}$ for $\gamma_2$. The total $C/T$ contribution then rises to 22 mJ mol$^{-1}$K$^{-2}$ if we assume one copy of each sheet. This is still much too small. Thus we assume multiple copies of each sheet. For example, 
$4\times \alpha$, $4\times \beta$, $2\times \gamma_1$ and $2\times\gamma_2$ gives 52 mJ mol$^{-1}$K$^{-2}$, in the range of the observed Sommerfeld coefficient, and the
total BZ occupation will be around 84\%. If, as seems more likely,  $\gamma_1$ and $\gamma_2$ are extremal orbits on the same cylindrical Fermi surface, then we can have $4\times\gamma$ instead of $2\times\gamma_1$ and $2\times\gamma_2$ with an average contribution of 9.1 mJ mol$^{-1}$K$^{-2}$ per sheet gives the same result. This scenario would account for all of the specific heat, but the BZ would be very packed. The only other possibility is that there are heavy orbits (\textit{i.e.} hot spots) on the Fermi surface sheets that we have observed, or else other heavy surfaces that we have not observed. A better estimate of $\gamma$ will be obtained once we have reliable band-structure calculations. 

\subsection{Comparisons to Previous Studies and Other Related Compounds}
Comparing our results to the earlier dHvA study on URhIn$_5$ by Matsumoto \textit{et al.} \cite{Matsumoto2014}, the frequency $\alpha$ and its splitting at higher angles were seen in both experiments. The angle dependence, which is indicative of an ellipsoidal Fermi surface, agrees well. In our experiment however, we also observed two larger Fermi surface sheets, $\gamma_1$ and $\gamma_2$, as well as one additional small spherical pocket
$\beta$. While the total contribution from the observed sheets to specific heat remains well under the experimental value
of $\gamma = 50 - 60$ mJ mol$^{-1}$K$^{-2}$, we can at least consider the possibility of some combination of these sheets as mentioned in the section above, that will make up the missing specific heat contribution. The observation of these larger sheets also suggests that URhIn$_5$ is very different from the non-magnetic, 5\textit{f}-itinerant semi-metal URhGa$_5$ which has only small pockets and small specific heat \cite{Ikeda2002,Matsumoto2014}.   

In Ref. \onlinecite{Matsumoto2014}, a comparison was made to a reference compound ThRhIn$_5$, which has the same crystal structure but does not have  5\textit{f} electrons. ThRhIn$_{5}$ has very large Fermi surfaces (up to 7 kT) and with strong angle dependence that would be expected for quasi-cylindrical surfaces \cite{Matsuda2007,Matsumoto2014}. These are considerably larger than our observations. However, as $\gamma_1$ and $\gamma_2$ are suggestive of being cylindrical surfaces, it may be that the
electronic structure of URhIn$_5$ is not as drastically different from ThRhIn$_5$ as previously thought \cite{Matsumoto2014}. 

UNiGa$_5$ is another antiferromagnet in this family, with a N\'eel temperature $T_N = 85.5$ K and a Sommerfeld coefficient $\gamma = 30$ mJ mol$^{-1}$K$^{-2}$ \cite{Tokiwa2001}. Its Fermi surface has also been mapped using dHvA. In URhIn$_5$, our $\alpha$ surface has a very different angular dependence from that of any of the ellipsoidal Fermi surfaces in UNiGa$_5$. The largest cylindrical Fermi surface in UNiGa$_5$, on the other hand, also has much smaller cross-section for field along the
\textit{c}-axis than our $\gamma_{1}$ and $\gamma_2$ surfaces. Whereas Tokiwa \textit{et al.} see clear evidences of cylindrical Fermi surfaces \cite{Tokiwa2001}, we were unable to follow our $\gamma_{1}$ and $\gamma_2$ frequencies far enough to confirm the 1/$\cos\theta$ dependence expected for a cylindrical Fermi surface, but on balance it seems very likely that they arise on a cylindrical surface. The effective
masses of UNiGa$_5$ at around 1.4 to 3.1 $m_e$ are, however, similar to those found for URhIn$_5$ \cite{Tokiwa2001} There are no calculations available for the antiferromagnetic
ground state of UNiGa$_5$, and the paramagnetic state calculation does not match the experimental results \cite{Tokiwa2001}.   

As mentioned in the introduction, another related compound to URhIn$_5$ is UIn$_3$. This compound is also an antiferromagnet and orders at $T_N = 88$ K \cite{Tokiwa2001a}. It has a Sommerfeld coefficient of $\gamma = 40$  mJ mol$^{-1}$K$^{-2}$. The Fermi surface has been mapped, and no cylindrical Fermi surfaces were found \cite{Tokiwa2001a}. The angle dependence does not resemble our findings for URhIn$_5$, and the effective masses of UIn$_3$  range from 9.8 to 33 $m_e$, which are significantly larger than
those of URhIn$_5$. However, as is the case of UNiGa$_5$, the lack of reliable band structure
calculations for URhIn$_5$ and UIn$_3$ \cite{Tokiwa2001a} makes these comparisons quite speculative. 

\section{Conclusion}
We remark that there are major discrepancies between the calculated 5\textit{f} all-itinerant Fermi surface and the experimentally obtained dHvA frequencies, and the cyclotron masses also differ. The present results cannot be fully explained by the paramagnetic calculation and it is therefore highly desirable that a detailed band structure calculation be carried out for the
antiferromagnetic ground state.

\section{Acknowledgements}
J. C. would like to acknowledge many helpful discussions with M. Divi\v{s}. This work was generously supported by NSERC and CIFAR of Canada, Canada Research chair, the Czech Science Foundation Grant No. P203/12/1201 and the Charles University project CA UK Grant No. 128317. 

\bibliography{PaperBib}
\end{document}